\documentclass[12pt, letterpaper]{article}
\usepackage[utf8]{inputenc}
\usepackage[margin=1in]{geometry}

\usepackage[dvipsnames,svgnames,x11names]{xcolor}
\usepackage[final]{changes}
\definechangesauthor[color=BrickRed]{DS}
\definechangesauthor[color=NavyBlue]{PS}

\usepackage{amsmath, amsthm, amssymb, calrsfs, wasysym, verbatim, bbm, color, graphics, geometry, indentfirst, amsfonts}

\usepackage[american]{babel}
\usepackage{csquotes}
\usepackage{natbib}
\usepackage{multirow}

\usepackage{doi}

\usepackage{array}
\usepackage{longtable}
\usepackage{booktabs}
\usepackage{tabularx}
\newcolumntype{E}{>{\raggedright\arraybackslash}X}
\newcolumntype{I}{>{\raggedleft\arraybackslash}X}
\usepackage{tabulary}

\usepackage{chngcntr}
\usepackage{titling}

\usepackage{graphicx}
\usepackage{subcaption}
\DeclareCaptionLabelSeparator{mytable}{.\\}
\providecommand{\keywords}[1]
{
  \small	
  \textbf{\textit{Keywords---}} #1
}

\usepackage{enumerate}

\theoremstyle{definition} 

\newtheorem*{remark*}{Remark} 

\usepackage{mathtools}

\usepackage{authblk}

\title{The Causal Effect of the Two-For-One Strategy in the National Basketball Association}

\author[1]{Prateek Sasan}
\author[2]{Daryl Swartzentruber}
\affil[1]{\protect\raggedright 
The Ohio State University, Department of Statistics, Columbus, Ohio, U.S.A., e-mail: sasan.1@buckeyemail.osu.edu}
\affil[2]{\protect\raggedright 
Centre College, Data Science and Mathematics Programs, Danville, KY, U.S.A., e-mail: daryl.swartzentruber@centre.edu}

\begin{document}

\maketitle

\abstract{This study evaluates the effectiveness of the “two-for-one” strategy in basketball by applying a causal inference framework to play-by-play data from the 2018-19 and 2021-22 National Basketball Association regular seasons. Incorporating factors such as player lineup, betting odds, and player ratings, we compute the average treatment effect and find that the two-for-one strategy has a positive impact on game outcomes, suggesting it can benefit teams when employed effectively. Additionally, we investigate potential heterogeneity in the strategy’s effectiveness using the causal forest framework, with tests indicating no significant variation across different contexts. These findings offer valuable insights into the tactical advantages of the two-for-one strategy in professional basketball.}

\makeatletter
\renewcommand{\abstractname}{Executive Summary}
\makeatother
\begin{abstract}
The two-for-one basketball strategy involves attempting to take a shot early in a possession in order to maximize the number of possessions at the end of a period. Based on our analysis of two seasons of data from the National Basketball Association, we estimate that teams that attempt a two-for-one strategy gain slightly more than half a point per attempt, on average, compared to teams that do not attempt a two-for-one strategy. We do not find that differences in certain relevant covariates lead to substantial differences in this point gain. Our analysis thus empirically validates the growing consensus of the effectiveness of the two-for-one strategy in the National Basketball Association.
\end{abstract}


\section{Introduction}
In recent years, the use of data and statistical 
methods to analyze sports strategies has gained 
significant traction, driven by the availability 
of detailed game data and advances in data 
science. In professional sports leagues like the 
National Basketball Association (NBA), decision-makers seek to optimize team 
performance through strategic adjustments 
informed by empirical analysis. Among these 
strategies, the “two-for-one” tactic, where a 
team attempts to gain an extra possession by 
timing shots near the end of a period, has 
become popular. However, despite its widespread 
use, little rigorous evidence exists on whether 
this strategy provides a measurable advantage. 
This study aims to address this gap by applying 
causal inference methods to determine the 
effectiveness of the two-for-one strategy.

An increasing body of research has applied causal inference methods to investigate the causal effects of various sports strategies. For example, \citet{gibbs2022causal} utilize causal inference to evaluate the impact of calling a timeout to disrupt an opposing team’s scoring run in NBA games. \citet{nakahara2023pitching} analyze pitching strategies in baseball through a causal inference framework, while \citet{yam2019lost} examine the decision-making process surrounding fourth-down plays in the NFL, aiming to quantify the effectiveness of these high-stakes decisions.  Additionally, \citet{wu2021contextual} explore the effect of crossing the ball in soccer, seeking to determine whether this strategy significantly influences scoring opportunities.

These studies represent a growing interest in the causal analysis of strategic decisions in sports, where the interplay of chance and skill presents unique challenges to identifying causality. By employing techniques that control for confounding variables and capitalize on observational data, such research provides insights that go beyond traditional statistical descriptions, aiming instead to uncover the causal impact of specific decisions. The increasing application of causal inference in sports analytics thus reflects a broader trend towards data-driven decision-making and evidence-based optimization of strategy across various sports contexts.

Causal  inference \citep{imbens2015causal} methods are essential tools for rigorously analyzing the effects of treatments or interventions in observational settings, particularly when randomized controlled trials (RCTs) are impractical, as is often the case in sports. In this study, we focus on estimating the average treatment effect (ATE) of the NBA’s two-for-one strategy. The ATE quantifies the average difference in outcomes (e.g., difference in score margin) between situations where the two-for-one strategy is employed and those where it is not. Estimating the ATE in observational data, however, requires addressing confounding variables—factors that influence both the treatment and the outcome, potentially biasing the results. To mitigate such bias, we apply augmented inverse probability weighting, a widely used method in causal inference that allows for more reliable estimation of treatment effects. This method involves estimating the probability (propensity score) that a given observation will receive the treatment (the two-for-one strategy) based on observed covariates (e.g., player lineup, time in the game). Each observation is then weighted by the inverse of its propensity score to create a pseudo-population with a similar distribution of covariates between treated and control groups. 

Beyond estimating the ATE, it is crucial to assess 
whether the treatment effect varies across different 
contexts or player configurations. To do this, we employ 
the causal  forest \citep{athey2019estimating, wager2018estimation} framework, a machine learning-based method that enables the 
estimation of heterogeneous treatment effects (HTE). 
Traditional ATE estimation assumes that the treatment 
effect is constant across all observations, but in real-world settings like the NBA, the effectiveness of a 
strategy is likely to vary based on game context, player 
lineups, and other factors. The causal forest framework 
addresses this by partitioning the data into subgroups 
using decision trees \citep{Hastie2009}. 
Each subgroup represents a 
distinct set of conditions where the treatment effect 
may differ. Causal forests construct many such trees and 
aggregate the resulting estimates to produce a flexible, 
non-parametric estimate of treatment effect 
heterogeneity. By identifying subgroups with distinct 
treatment effects, causal forests allow us to uncover 
whether the two-for-one strategy is more or less 
effective in different situations, such as when certain 
players are on the court or at different points in 
a game.

In addition to estimating treatment effect 
heterogeneity, we apply the concept of rank 
average treatment effect  (RATE) \citep{yadlowsky2024evaluating} to better 
understand the strategy’s impact. RATE works by 
ranking observations based on their predicted 
treatment effects, identifying which game contexts 
or player configurations are most likely to 
benefit from the two-for-one strategy. The RATE 
method quantifies the expected impact on the top
ranked subgroups and compares these with the 
average effect across all observations. This 
ranking approach is particularly valuable in 
practice, as it highlights the specific conditions 
under which the two-for-one strategy is most 
advantageous, offering actionable insights for 
coaches and analysts looking to optimize in-game 
decisions.

The structure of this paper is as follows. Section 2 presents definitions and background on the two-for-one strategy and details the data set used in our analysis. Section 3 outlines the causal methodology employed to assess the impact of this strategy. Section 4 presents the results obtained from analyzing two seasons of NBA data, highlighting key findings on the effectiveness of the two-for-one approach. Section 5 concludes with a brief discussion.

\section{Definitions and Data}
In this section, we provide a detailed explanation of the two-for-one (TFO) strategy in the NBA, including a mathematical definition of the strategy and a description of the data sources used in this study.

The TFO \citep{feng_2015, fischer_2015} strategy in the NBA is a time-management tactic that aims to maximize a team's scoring 
opportunities by securing two offensive possessions 
at the end of a period, while limiting the 
opponent to just one. This strategy is commonly employed 
in the last 30–40 seconds of a period, especially 
during close games, where maximizing scoring chances can 
be crucial. The TFO strategy is primarily 
focused on managing the game clock to optimize 
possessions and increase scoring potential before the 
period ends. To implement the TFO, a team aims to take a shot 
with a certain amount of time remaining in the period so that even if 
the opponent uses a full possession, the team can secure 
the ball again before time expires. 

\added{In order to attempt a TFO, a team may be required to take a shot early in the possession, regardless of the quality of shot. This runs counter to the common strategy in basketball of progressing through an offensive set until an opportunity for a quality shot is realized. Thus the potential down side of attempting a TFO is that even if a team ends up with more possessions at the end of the period than their opponent, they may not ultimately benefit because of poorer scoring opportunities.}

\subsection{Definitions} \label{definitions}
\replaced{To see if the potential gains of a TFO strategy outweigh the potential losses we need to determine a causal effect, and for that}{
In order to determine the causal effect of the TFO strategy in the NBA,} we need a well-defined 
treatment. That is, we need to be able to classify 
scenarios during an NBA game as those in which a team 
enacted the TFO strategy, and those in 
which they did not. This is not a trivial matter. Teams 
do not announce that they are attempting a TFO, and it 
is unclear how attune players are to the game and shot 
clock during any particular play. Thus we rely on a 
combination of time and play outcomes in order to 
identify TFO \emph{opportunities}. These are scenarios 
when a team comes into possession of the ball with 
enough time left in the period such that taking a shot 
or initiating offense relatively early in the shot clock 
leads to a increased chance to get an extra possession, 
while waiting until the end of the shot clock to do 
these things increases the chance the opposing team will 
have the last possession. We then categorize those TFO opportunities 
into \emph{attempts} (treatment group) and \emph{non-attempts} (control group).

Several important characteristics of NBA game play inform these definitions. NBA games are played in four 12-minute periods. Possession at the beginning of the game is determined by a jump ball, the result of which also determines which team gets possession at the start of subsequent periods. The shot clock in the NBA is 24 seconds, meaning that from the time a team gains possession of the ball after a turnover, defensive rebound, or made basket by the opposing team, the team has 24 seconds in which to attempt a shot. If a team rebounds their own missed shot the shot clock is reset to 14 seconds instead of 24. If a team gains possession with less than 24 seconds left in the period, the shot clock is turned off and the team in possession can attempt to have the last possession of the period by waiting until the very end to take a shot. Therefore, if a team wants to improve its chances of gaining the last possession of the period, it should hope that the opposing team gains possession with more than 24 seconds left. 

However, it can take up to a few seconds for possession to change, particularly after a missed shot. As an example, suppose a player on Team A takes a shot with 6 seconds left in the period. It takes one second for the ball to get from the player's hand to the rim, another second to bounce on the rim a couple of times before coming to the ground, and a third second before a player on Team B possesses it, now with only 3 seconds left in the period. Furthermore, if there is too little time left in the period when a team gains possession, that team has a very small chance of scoring any points. 

With these characteristics in mind, we set a limit of 28 
seconds left in the period, and say that if a team 
takes a shot after this point and the possession 
changes, it is unlikely that they will get a quality 
possession at the end of the period. Working backwards, 
then, a team needs to gain possession of the ball with 
enough time before the 28 seconds mark in order to 
potentially attempt a reasonable shot by this point. We 
set this lower limit to be 35 seconds. We choose 43 seconds as an upper limit on when a team can gain possession to exclude those scenarios where a team can get to this optimal shot time simply by waiting until the shot clock for their possession is close to 0. Thus we define a TFO opportunity as one in which a team comes into possession of the ball with between 35 and 43 seconds left in the period. 

If a team presented with a TFO opportunity takes a shot or is fouled while taking a shot with at least 28 seconds to go in the period, we consider that a clear TFO attempt. Other play results are harder to categorize. We also consider an opportunity an attempt if the team is fouled with at least 28 seconds to go in the period, with the rationale being that fouls more often occur when a team is trying to initiate offense. On the other hand, we consider an opportunity to be a non-attempt if the team maintains possession until there are less than 28 seconds left in the period without taking a shot or being fouled. 

We remove all other play outcomes that occur during a TFO opportunity from consideration rather than defining them as either an attempt or a non-attempt. The most common of these is a turnover. Turnovers can happen while a team is initiating offense, such as a charge foul, or just trying to hold the ball until the end of the shot clock, such as a steal that happens far away from the basket. Due to the lack of information on turnovers available in play-by-play data, we choose to remove them from our analysis. The same is true for other, less common plays such as a jump ball, a defensive three-second violation, and a kicked ball violation. Figure 1 provides a visual description of our treatment and control groups.

\begin{figure}[t!]
\centering
\includegraphics[width=\textwidth]{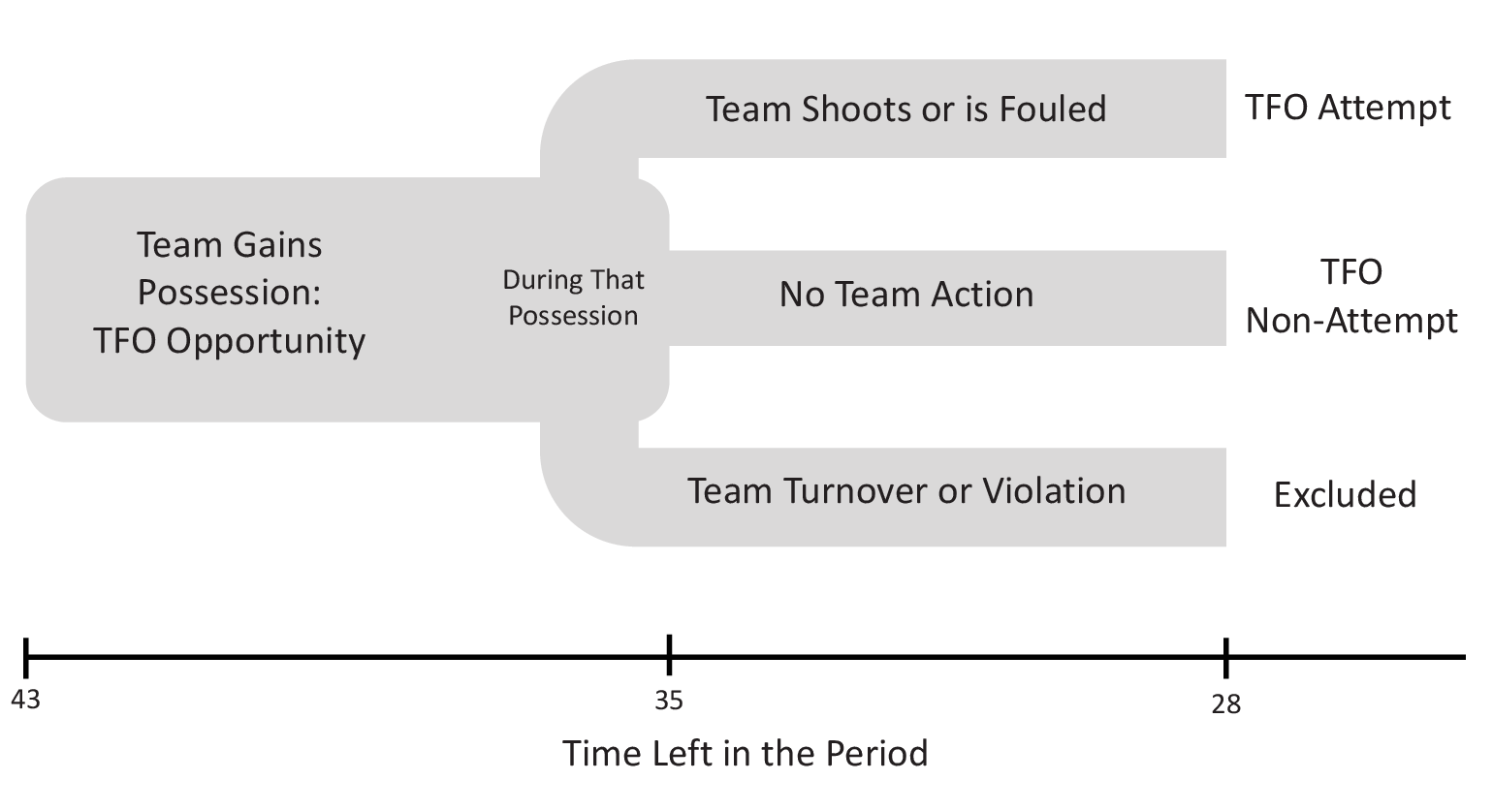} 
\caption{Diagram of the treatment definition.}
\label{fig:treat}
\end{figure}

We believe this to be a reasonable treatment definition for the TFO strategy. However, it also has limitations. One of these is that the timing cutoffs, while well motivated, are nevertheless somewhat arbitrary. Thus in Section \ref{treatsens} we present a sensitivity analysis in which we adjust these cutoffs slightly to explore the robustness of our estimates to changes in the definition. 

A second limitation is that, as mentioned earlier, there is no way to know the intent of the coaches or the players on the floor with respect to the TFO strategy, which forces us to rely on event data for our treatment definition. There are certainly scenarios when a player intends to implement a TFO strategy but is not part of our treatment group because, for example, they lost control of the ball while going up for a shot and it is recorded as a turnover. The distinction from the clinical trials literature between intent-to-treat (ITT) and per-protocol (PP) analyses is relevant here. In the former case subjects are analyzed based on the original intended treatment, where in the latter they are analyzed based on the treatment they actually receive. \citet{MOLEROCALAFELL2024111457} discuss the strengths and limitations of these analyses in relation to whether a study is more explanatory, meaning executed under more ideal experimental conditions, or pragmatic, executed under more real world conditions. Certainly our study is more pragmatic by their definitions, and thus they would recommend an ITT analysis with perhaps a PP analysis as a complement. However, in this framework we do not know the original intent and thus our approach is more similar to that of a PP analysis because an ITT analysis is not possible. However, \citet{MOLEROCALAFELL2024111457} mention that employing advanced statistical methods such as the inverse probability weighting we use in our analysis can help enhance the validity of PP estimates.

We define our response variable as the Post Opportunity Difference (POD), which measures the difference in score differential from the time of the TFO opportunity to the end of the period. A positive POD indicates that the team 
increased its score 
margin by the end of the period. Conversely, a negative 
POD implies that the team’s score margin decreased 
after the TFO opportunity. Thus if the TFO strategy is effective the POD would be higher for teams that attempt a TFO than those who do not.

\subsection{Data} \label{data}
The main source of data we use is play-by-play data from the NBA. This is a list of all the different plays that took place in an NBA game as well as supplemental information such as the players involved in the play, the score of the game at the time of the play, the period the play took place in, and the amount of time left in that period. We obtain play-by-play data from the nbastatR package \citep{nbastatR}, which interfaces with the NBA Stats API and other reference sources to give a variety of data and summary statistics at the player, team, game, and season level. This includes play-by-play data dating back to the 1996-1997 season, the first year the NBA kept such data. For our analysis we focus on two recent seasons, that of 2018-2019 and 2021-2022. The 2018-2019 season was the first regular season under a modified set of shot clock rules as well as the last regular season before the Covid-19 pandemic altered the NBA schedule, and the 2021-2022 season was the first full season after the pandemic, making these natural seasons to consider in order to explore the overall effect of the TFO strategy as well as possible heterogeneity across multiple seasons. 

Using this play-by-play data, we identify all TFO opportunitys in the first three periods of each NBA regular season game. We choose to exclude the fourth periods \added{and overtime} because the behavior of NBA games is often much different at the end of those periods. If the game is not close, teams are no longer necessarily worried about maximizing their possessions because of the feeling that nothing they do will affect the outcome. If the game is very close, teams tend to utilize a strategy of intentionally fouling in order to increase the number of possessions in the game. We believe that excluding these scenarios allows us to better isolate the treatment\added{, and we present a numerical justification for this choice in Section \ref{perperiod}}. We also exclude TFO opportunities in which the team has already had an earlier TFO opportunity in the same period. For the 2018-2019 season, there were a total of 1567 TFO opportunities, of which 1036 were classified as attempts and the remaining 529 classified as non-attempts. For the 2021-2022 season there were a total of 1479 TFO opportunities, of which 950 were classified as attempts and the remaining 462 classified as non-attempts. Each NBA season contains 1230 regular season games, so there was an average between 1 and 2 TFO opportunities per game based on our criteria.

To enhance our analysis with relevant covariates, we draw from several supplemental data sources. First, to gauge the relative strength of the teams at the time of each game, we incorporate betting odds, which indicate the favored team and the expected point spread. These odds effectively summarize expert assessments of each team's strength. Additionally, to measure the projected pace of play, we include betting totals, which provide a prediction of the total points expected to be scored in a game. Both the betting odds and totals are sourced from the Sportsbook Reviews archive \citep{nbaodds}.

For player-specific information, we use individual player ratings from the NBA 2K video game series \citep{nba2k19, nba2k22} which provide relative strength assessments for each player. While not perfect, these ratings are a reasonable proxy for the overall ability of players. In addition to previous years' statistics, ratings are informed by expert projections, particularly for players new to the league \citep{irving2024nba2k, ronnie2k2024ratings}. We obtain starting lineups from BasketballReference.com  \citep{basketballreference} and use this information, combined with play-by-play data, to determine which players are on the court at any given time. These covariates allow us to account for variations in team and player strength as well as game tempo, enhancing the robustness of our analysis. Table \ref{tab:variable_names_def} provides the name and definitions of the variables we use.

\begin{table}[t]
\centering
\renewcommand{\arraystretch}{1.2} 
\begin{tabularx}{\textwidth}{|c|X|}
\hline
\textbf{Variable Name} & \textbf{Variable Definition} \\
\hline
Period & Period of the game; data is restricted to the first three periods. \\
Time Left & Remaining time in the current period, measured in seconds. \\
Score Margin & Difference between the scores of the two teams.   \\
Max Rating & Maximum player rating among players on the court for the team at that moment. \\
Max Rating Opposition & Maximum player rating among players on the court for the opposing team at that moment. \\
Mean Rating & Average player rating of the players on the court for the team at that moment. \\
Mean Rating Opposition &  Average player rating of the players on the court for the opposing team at that moment. \\
Spread &  The projected difference in final score for the two teams, set by oddsmakers. \\
Total Score & The projected sum of the scores of the two teams, set by oddsmakers. \\
\hline
\end{tabularx}
\caption{List of variables used in the propensity score model along with their definitions.}
\label{tab:variable_names_def}
\end{table}

Combining these covariates and the play-by-play data into a single usable data set requires a substantial amount of data cleaning. The play-by-play data includes substitutions but not a list of the current players on the floor, so we have to input the starting lineups and then create variables that track the current lineup on the floor. Player names are also represented differently in the various sources which complicates data merging. Finally, the play-by-play data provides narrative descriptions such as the type of shot, which is a challenge when categorizing plays. As an example, in Table \ref{tab:playbyplay} described in the section below, the phrases \emph{jump shot}, \emph{finger roll layup}, and \emph{dunk} all indicate a shot has been attempted.

\subsection{Example} \label{example}
To illustrate the definitions and data described above, consider an example from a game played between the Phoenix Suns and the Golden State Warriors on October 22, 2018. Table \ref{tab:playbyplay} shows several rows of the play-by-play data with a subset of relevant columns. In the first period of the game the Suns gained possession with 42 seconds left after a Warriors turnover, giving the Suns a TFO opportunity. However, they did not take a shot early in the shot clock. Rather, they committed a turnover with 25 seconds left. By our definition, this is a non-attempt. The Warriors gained possession with the turnover and made a 3 point shot, after which there was no more scoring in the period. So from the time of the TFO opportunity until the end of the period the Suns scored 0 points and the Warriors scored 3 points, for a POD of -3.

With 37 seconds left in the second period of the same game, the Warriors gained possession after a made shot by the Suns, giving the Warriors a TFO opportunity. With 31 seconds left Kevin Durant made a dunk for the Warriors, and we consider this shot attempt with more than 28 seconds to go in the period a TFO attempt. The Suns gained possession at that point and committed a turnover with 12 seconds left. Thus the Warriors gained a final possession, and Quinn Cook made a three-point shot with one second left, which did not allow the Suns to take another shot. From the time of the TFO opportunity until the end of the period, the Warriors scored 5 points and the Suns scored 0 points, for a POD of 5.

\begin{table}
\centering
\normalsize
\begin{tabularx}{\textwidth}{llllll}
Pr. & Time & DescriptionPlayHome & DescriptionPlayVisitor & ScoreH & ScoreV \\ 
\midrule\addlinespace[2.5pt]
1 & 0:42 & Curry Out of Bounds -& NA & NA & NA \\ 
&& Bad Pass Turnover &&&\\
&& Turnover (P3.T5)&&&\\
1 & 0:25 & NA & Booker Discontinue   & NA & NA \\ 
&&&Dribble Turnover (P2.T3)&&\\
1 & 0:25 & SUB: Cook FOR Looney & NA & NA & NA \\ 
1 & 0:25 & SUB: Thompson FOR  & NA & NA & NA \\ 
&&McKinnie&&&\\
1 & 0:07 & Jerebko 27' 3PT Step   & NA & 32 & 23 \\
&&Back Jump Shot (3 PTS) &&&\\
&&(Thompson 2 AST)&&&\\
1 & 0:03 & Iguodala STEAL (1 STL) & Booker Lost Ball Turnover & NA & NA \\ 
&&& (P3.T4)&&\\
1 & 0:01 & MISS Curry 55' 3PT & NA & NA & NA \\ 
&&Jump Shot &&&\\
1 & 0:00 & WARRIORS Rebound & NA & NA & NA \\ 
\hline
2 & 0:37 & NA & Booker 4' Driving Finger  & 65 & 47 \\ 
&&&Roll Layup (14 PTS)&&\\
2 & 0:31 & Durant 3' Driving Dunk & NA & 67 & 47 \\ 
&&&(15 PTS) &&\\
2 & 0:12 & NA & Booker Out of Bounds  & NA & NA \\ 
&&&Lost Ball Turnover (P5.T9)&&\\
2 & 0:12 & SUB: Cook FOR Jones & NA & NA & NA \\ 
2 & 0:12 & NA & SUB: Bridges FOR & NA & NA \\ 
&&&Anderson &&\\
2 & 0:01 & Cook 24' 3PT Pullup & NA & 70 & 47 \\ 
&&Jump Shot (4 PTS) &&&\\
&&(Thompson 3 AST)&&&\\
\end{tabularx}
\normalsize
\caption{Section of play-by-play data obtained from the nbastatR package. The variables shown are the period of play, the time left in that period, descriptions of the play for both the home and visitor teams, and score for both the home and visitor teams. Variable names have been altered.}
\label{tab:playbyplay}
\end{table}

\section{Methodology}
Using the definitions from the previous section, we define the treatment $W$ for TFO opportunity $i$ as $W_i=1$ if the TFO is attempted and $W_i=0$ if the TFO is not attempted. We let $Y_i$ be the POD. We employ the potential outcomes framework \citep{Neyman1923OnTheApplication, rubin1974} in order to define the primary causal effect of interest. Let $Y_i(1)$, $Y_i(0)$ be the potential outcomes under treatment and control, respectively. That is, $Y_i(1)$ is the POD a team would get if it attempts a TFO at opportunity $i$, and $Y_i(0)$ is the POD a team would get if it does not attempt a TFO at opportunity $i$. A natural estimand of interest is the average treatment effect (ATE):
\begin{equation}
    ATE=E[Y_i(1)-Y_i(0)].
\end{equation}
The ATE quantifies how much larger the POD would be, on average, if NBA teams presented with a TFO opportunity always made a TFO attempt rather than never made a TFO attempt.

\replaced{The fundamental problem of causal inference is that} {Unfortunately,} we do not observe both $Y_i(1)$ and $Y_i(0)$ for any particular observation. \deleted{For each TFO opportunity, it is impossible to know what would have happened if the team had chosen a different path. This is known as the fundamental problem of causal inference.} If the scenarios where a team attempted a TFO were the same, on average, in all meaningful ways to the scenarios where a team did not attempt a TFO, an assumption known as exchangeability, then we could estimate the ATE from the data as $\hat{E}(Y_i|W_i=1)-\hat{E}(Y_i|W_i=0)$. Unfortunately that assumption is not reasonable in this situation, as potentially confounding variables such as the score margin and the time left in the period are likely to influence a team's decision.

The methods of causal inference have been developed for precisely such circumstances, when a causal effect estimate is desired for observational data where exchangeability does not hold. Many of these methods rely on a common set of assumptions, namely consistency, no interference, conditional exchangeability, and positivity. Below we discuss these assumptions and why they are reasonably satisfied for our data, followed by a description of the specific methods we will be using to estimate treatment effects.

\subsection{Causal Assumptions} \label{assumptions}
The \emph{consistency} assumption implies that there is a well-defined treatment with no hidden variations. In our setting we recognize that not all TFO attempts look the same, as highlighted in the previous section. Some end in a shot while others end in a foul. In some cases the team that had the TFO attempt gets the ball back, and in other cases it does not. However, because these differences are a part of the definition of our treatment, they are not hidden and do not form an egregious violation of the consistency assumption. 

The \emph{no interference} assumption says that the outcome for a particular unit does not depend on the treatment assignment of any other unit. Mathematically, this means that for any individual 
$i$, 
\begin{equation}
 Y_i(W_i, W_{-i}) = Y_i(W_i)   
\end{equation}
where $W_{-i}$ represents the treatment vector excluding 
$i$. This assumption would be violated in our setting if 
the success of a particular team's TFO attempt was affected by 
whether another TFO opportunity in the study was an attempt or a non-attempt. 
One potential source of interference could come from the possibility of multiple TFO opportunities in the same period. For example, suppose a team gains possession with 43 seconds left and makes a shot with 36 seconds left, after which the opposing team gains possession and makes a shot with 29 seconds left. Then both teams have attempted a TFO. However, we do not believe that the success of the TFO attempt by the team gaining possession with 36 seconds left in this scenario depends on the fact that the first team is also attempting a TFO. Had the first team gained possession with 50 seconds left instead of 43, we would not consider this a TFO attempt but the second team gets the ball in an practically identical situation as the original scenario. Thus we have left these instances in our data set.

A slightly more complicated scenario occurs if the same team has back-to-back TFO opportunities. This may happen, for example, if a team gains possession with 43 seconds left, misses a shot with 36 seconds left, but gets an offensive rebound, thus earning another possession. Often in this case the player that gets the rebound is able to take a quick shot close to the goal which has a higher than average probability of going in or drawing a foul. So it is plausible that these TFO attempts may lead to a higher POD because they follow another TFO attempt from the same team. Thus we have removed these instances from our data set.

Other than units occuring in the same period, it is difficult to imagine a scenario where there would be a serious violation of this assumption. It is possible 
due to the season-long nature of the data that if a 
particular team has a large number of TFO attempts relative to 
non-attempts early in the year, the players may practice last-second shots more because they have been experienced 
them more in the course of play, and thus they could 
improve the outcome of the TFO attempts later in the year. 
However, players already practice these types of shots 
because of their importance to game play, and it seems 
unlikely that the amount of time they spend practicing 
such shots substantially depends on how often they 
attempt a TFO strategy. Thus the no interference 
assumption seems to reasonably hold for our setting.

\emph{Conditional exchangeability} implies that, given a set of covariates $X$, the treatment assignment $W$ is independent of the potential outcomes: 
\begin{equation}
    Y_i(0), Y_i(1) \perp W_i \mid X_i.
\end{equation}
This allows us to assume that treated and untreated individuals with the same values of 
$X$ have comparable distributions of potential outcomes, thus enabling causal comparisons. Suppose our set of covariates consists of the score margin and the time left in the period. Then conditional exchangeability would say, for example, that TFO opportunities where a team is down by 4 with 40 seconds left in the period are the same, on average, for both the set of attempts and non-attempts. 

One way to examine the reasonableness of this assumption is to model the relationships between treatment, response, and covariates in a directed acyclic graph (DAG). In these diagrams, an arrow from one node to another represents a causal relationship between those variables. For a DAG to be complete, it must contain all common causes of any variables in the graph. Under this assumption, we can use a quick visual check called the back-door criteria to determine whether a set of covariates satisfies the conditional exchangeability assumption. See \citet{Pearl_2009} for more information on the use of DAGs in causal inference.

We propose the DAG given in Figure 2 as a model for our causal structure, using the covariates in Table \ref{tab:variable_names_def}. Note that we combine the ratings variables into a single node for the sake of simplicity and because we think they would have the same relationship structure with the other variables. We think that each of these covariates could reasonably affect both the treatment and response variables. Teams may be more likely to attempt a TFO if there is more time left in the period, and this also could possibly allow for more possessions after the TFO opportunity, which could increase the POD. There spread and the score margin represent the expected and current point difference between the teams, and it is possible that teams that are behind or are expected to be behind may feel more of a need to maximize possessions. At the same time these teams may score less on average per possession than those winning or expected to win, affecting the POD. The total score is in some ways a proxy for pace of play. Teams that play faster may be more likely to attempt a TFO than a more methodical team, and a faster pace can lead to more possessions and thus affect the POD. The quality of the players on the floor certainly impacts the scoring at the end of the period, but it may also impact the likelihood of a TFO attempt if, for example, a team playing its best players is able to execute well against an opponent playing weaker players. Finally, we allow for the existence of period specific effects as team strategies may well be different during different parts of the game.

\begin{figure}[t!]
\centering
\includegraphics[width=.8\textwidth]{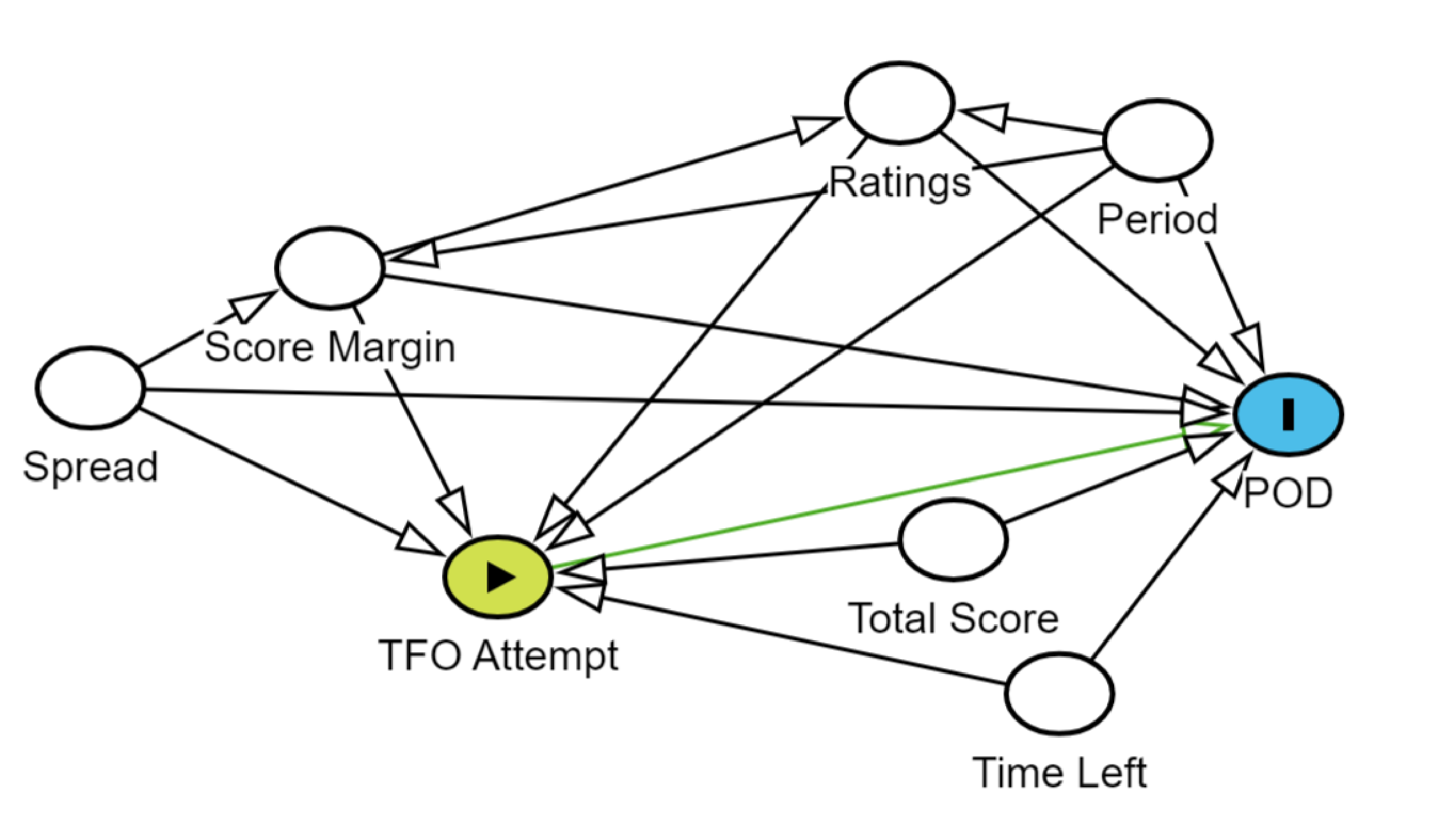} 
\caption{DAG of our proposed causal structure, created using the \textit{DAGitty} web application \citep{dagitty}}
\label{fig:dag}
\end{figure}

In addition to these confounding relationships, there are several relationships between covariates included in the DAG. For example, some coaches have set substitution patterns, and so the period of the game affects the ratings of the players on the floor. Alternatively, if a team is ahead by a large amount, a coach may change this pattern and rest his best players more than in a close game. It is important that none of these extra relationships lead to a cycle in our graph, which is not allowed in a DAG and would violate temporal restrictions of causality. Letting $X$ be the set of covariates listed in our DAG, the back-door criteria and thus conditional exchangeability is satisfied. However, while we believe that we have included the most important sources of confounding, there is the possibility of unmeasured confounding in any causal analysis. In Section \ref{ate_sens} we examine the sensitivity of our estimates to unmeasured confounding. It is also appropriate to check conditional exchangeability by looking at the balance of the covariates in the groups created by the methods discussed below, which we do in Section \ref{ass_checks}.

The final assumption necessary is the \emph{positivity} assumption, which says that each TFO opportunity must have a positive probability, conditional on our set of covariates, of being in either the treatment or control group i.e.,
\begin{equation}
    0 < P(W_i=w_i \mid X_i) < 1, \quad w_i = 0,1
\end{equation}
There is a known trade-off between positivity and 
conditional exchangeability. If you condition on too 
many covariates, you may satisfy the latter but not the 
former. If you do not condition on enough covariates, 
the opposite may be true. We feel that we have found a 
good balance between the two by carefully selecting a 
small but rich set of covariates. We perform a visual check of the positivity assumption in Section \ref{ate_results}.

\subsection{Causal Methods for Average Treatment Effects} \label{ate_methods}

Many causal inference methods rely on the concept of a propensity score \citep{rosenbaum1983central}, which is defined as the probability of receiving treatment given baseline covariates \added{$X$}. For a binary treatment $W \in \{0,1\}$, the propensity score denoted by $e(X)$ is given by 
\begin{equation}
    e(X) = P(W=1 \mid X).
\end{equation}
\replaced{Propensity scores are generally unknown quantities but can be 
estimated using logistic regression or machine learning 
methods.}{This scalar summary of multiple covariates facilitates 
covariate balance between treated and control groups, 
aiming to replicate the distributional balance of a 
randomized experiment.} 

Propensity scores are used to estimate treatment effects in several ways. Inverse probability weighting (IPW) is a class of estimators that assigns weights to individuals to create a pseudo-population where the distribution of covariates is balanced between treatment groups. \deleted{For an individual $i$ with covariates $X_i$ when the ATE is the estimand of interest, weights are computed as $\frac{W_i}{\hat{e}(X_i)} + \frac{1-W_i}{1-\hat{e}(X_i)}$ where $\hat{e}(X_i)$ is the estimated propensity score.} \replaced{These weights are}{This weight is} larger for individuals in groups with a lower 
probability of treatment assignment, adjusting for 
selection bias by up-weighting underrepresented 
individuals \citep{hirano2003efficient}. If conditional exchangeability holds in the true population, then marginal exchangeability holds in the pseudo-population. Thus the ATE can then be estimated by the weighted difference in outcomes,
\begin{equation}
    \hat{ATE}_{IPW} = \frac{1}{n} \sum_{i=1}^n \Big( \frac{W_iY_i}{\hat{e}(X_i)} - \frac{(1-W_i)Y_i}{1-\hat{e}(X_i)} \Big),
\end{equation}
\added{where $\hat{e}(X_i)$ is the estimated propensity score.}

Note that the set of covariates $X_i$ contains information about both the treatment assignment and the outcome. This motivates an alternative approach known as augmented inverse probability weighting (AIPW)  \citep{Glynn_Quinn_2010}. In addition to estimating propensity scores, this method estimates the outcome under both treatment and control, denoted $\hat{E}(Y_i|W_i=1,X_i)$ and $\hat{E}(Y_i|W_i=0,X_i)$, respectively. The IPW estimator is then adjusted by a term involving the weighted average of these two estimators:

\begin{multline}
        \hat{ATE}_{AIPW} = \frac{1}{n} \sum_{i=1}^n \left\{ \Big( \frac{W_iY_i}{\hat{e}(X_i)} - \frac{(1-W_i)Y_i}{1-\hat{e}(X_i)} \Big) - \frac{W_i-\hat{e}(X_i)}{\hat{e}(X_i)(1-\hat{e}(X_i))} \right. \\
        \left. [1-\hat{e}(X_i)\hat{E}(Y_i|W_i=1,X_i)+\hat{e}(X_i)\hat{E}(Y_i|W_i=0,X_i)] \right\} .
\end{multline}

This AIPW estimator is doubly-robust in that it is consistent for the ATE whenever either the propensity score model or the outcome model is correctly specified. It also requires the conditional exchangeability assumption mentioned above.

\subsection{Causal Methods for Heterogenous Treatment Effects}

\replaced{
While the ATE summarizes the overall impact of executing a TFO, it is natural to expect that its effect varies across different possession contexts. For example, some teams are naturally more effective at generating quick, quality scoring chances under time pressure, whereas others tend to rely on slower, more structured offensive sets that may yield lower-quality shots when rushed. Likewise, a TFO taken when a team holds a comfortable lead may influence subsequent decision-making differently than one taken in a highly competitive situation. The defensive characteristics of the opponent also matter: facing a unit that is slower to adjust to rapid possessions may create more favorable scoring conditions than facing a unit that is disciplined and well-positioned. These contextual differences may naturally lead to heterogeneous treatment effects. Using causal forests to detect this heterogeneity helps clarify when and in what types of game states the 2-for-1 is most advantageous, complementing the ATE without affecting its interpretation or underlying assumptions.
}{The ATE provides 
insight into the overall causal effect for a 
population. However, treatment effects often 
vary across individuals within the population.
The presence of heterogeneity in treatment 
effects can obscure important subpopulation 
dynamics if only the ATE is considered. 
Identifying and analyzing these differences 
allows for more tailored interventions, 
improving both individual and aggregate 
outcomes. Furthermore, examining treatment 
effect heterogeneity can enhance our 
understanding of how covariates interact with 
the treatment, providing deeper insights into 
causal mechanisms. This motivates the need for 
methodologies that move beyond the ATE to 
uncover nuanced variations in causal effects 
across different subgroups.}

The Conditional Average Treatment Effect (CATE) \citep{athey2016recursive} is a statistical measure used 
to capture the heterogeneity in treatment effects across 
different subgroups within a population. It extends the 
concept of the ATE, which 
calculates the expected difference in outcomes due to a 
treatment across the entire population, by conditioning 
on specific covariates $X$. This approach is 
particularly useful in causal inference to identify how 
treatment efficacy varies across different individuals 
or groups, allowing for more personalized interventions.

Mathematically, the CATE is defined as:
\begin{equation}
    CATE = E[Y_i(1)-Y_i(0) \mid X=x].
\end{equation}
By conditioning on 
$X=x$, CATE captures the expected treatment effect for 
individuals with characteristics $X=x$. Assuming 
conditional exchangeability and consistency, we can estimate CATE from observational or 
experimental data by modeling the outcome response 
surface.

\replaced{Classical approaches to heterogeneous treatment effect (HTE) estimation typically rely on subgroup analysis defined by observed covariates. Regression-based methods require pre-specifying interactions or assume homogeneity within subgroups, which can limit their ability to capture complex patterns. Recent work therefore focuses on flexible machine learning methods that discover heterogeneity in a data-driven manner. For example, \citet{athey2016recursive} introduced causal trees to identify subpopulations with different treatment effects, and \citet{wager2018estimation} extended this framework to causal forests, providing individual-level treatment effect estimates along with measures of uncertainty. Propensity score–based approaches \citet{rosenbaum1983central} can also be adapted for HTE analysis, though subclassification may struggle in high dimensions; methods such as BART \citep{hill2011bayesian}  and other flexible learners are commonly used to mitigate these issues.
In our analysis, we estimate HTEs using causal forests.}{Classical approaches to heterogeneous treatment effect (HTE) estimation often rely on 
subgroup analysis, where subgroups are defined by 
observed covariates. However, traditional regression-based methods may be limited by the need to pre-specify 
interactions or assume homogeneity within subgroups. 
Recent advances have thus focused on machine learning 
techniques, which allow for more flexible, data-driven 
discovery of treatment effect heterogeneity. For 
instance, Athey \& Imbens (2016) introduced causal 
trees for identifying subpopulations with distinct 
treatment effects, while Wager
\& Athey (2018) extended this to causal forests, a method based on 
random forests that provides both individual-level 
treatment effect estimates and uncertainty 
quantification.}

\deleted{Propensity score-based methods are also adapted for HTE 
analysis, such as subclassification on propensity scores 
within stratified groups (Rosenbaum \& Rubin 1983). 
However, propensity-based models assume covariate 
balance is achieved within each stratum, which may not 
hold in practice for high-dimensional data. To address 
this, flexible machine learning models, including 
generalized additive models and Bayesian Additive 
Regression Trees (BART), are commonly used to improve 
balance across strata and yield robust HTE estimates (Hill 2011). In our analysis, we use the causal forest framework to 
compute heterogenous treatment effects.}

\subsubsection{Causal Forests}

Causal forests  \citep{athey_tibs_2019, athey2019estimating,wager2018estimation} are flexible, nonparametric methods for estimating conditional average treatment effects (CATEs) and exploring treatment effect heterogeneity across covariates. They extend the random forest framework \citep{breiman2001random} by adapting both the splitting rules and estimation strategies to the causal inference setting. The procedure begins with an orthogonalization step, inspired by the double machine learning framework \citep{chernozhukov2018generic} in which both the outcome and treatment variables are residualized with respect to observed covariates. This removes variation explained by potential confounders, ensuring that differences in estimated treatment effects are not driven by baseline imbalances. Residualized variables are then used to grow decision trees where the splitting criterion is chosen to maximize heterogeneity in treatment effects between child nodes, rather than minimizing prediction error of the outcome. This design allows the trees to uncover complex, nonlinear patterns in how treatment effects vary across the population.

To prevent overfitting and obtain valid inference, causal forests employ an “honest” estimation strategy: the data are split into two parts, with one subsample used to determine the structure of the tree (i.e., the splits) and the other used to estimate treatment effects within the resulting leaves. This honest splitting ensures that the estimated treatment effects are asymptotically unbiased. Many such trees are grown on bootstrap samples of the data, and their predictions are aggregated to form the final CATE estimate for each observation, reducing variance and improving stability. \replaced{Because orthogonalization reduces confounding and the splitting rule targets heterogeneity, causal forests can uncover treatment effect patterns that parametric models may miss. They also handle high-dimensional covariates and complex interactions while providing valid inference. These properties make causal forests well-suited for identifying subpopulations with differing treatment responses.}{Because the orthogonalization step adjusts for confounding and the splitting rule focuses on heterogeneity, causal forests can detect treatment effect patterns that traditional parametric models would miss without strong functional form assumptions. Furthermore, the method naturally accommodates high-dimensional covariate spaces and complex interaction structures, while still providing valid statistical inference under reasonable assumptions. These features make causal forests a powerful tool for identifying subpopulations that benefit more (or less) from an intervention, guiding both policy targeting and scientific understanding.}

\section{Results}
In this section, we present the results of estimating 
the overall causal effect of the TFO strategy using AIPW. We then examine potential heterogeneity 
in treatment effects by employing the causal forest 
framework. We perform all calculations in R \citep{R}.
We implement the AIPW estimator using the CausalGAM package \citep{CausalGAM}. We fit the propensity score model using a generalized additive model for a binomial outcome with a probit link function. We fit the outcome models using a generalized additive model for a conditionally Gaussian outcome with an identity link function. We use an empirical sandwich estimator of the sampling variance from \citet{lunceford} to conduct inference.  

\subsection{Average Treatment Effect} \label{ate_results}

We compute the ATE separately for each season, 
as well as for the combined dataset that 
includes both seasons, using the methods described in Section \ref{ate_methods}. Estimating the ATE 
separately provides insight into the stability 
of the treatment effect across seasons, while 
combining the data allows for a larger sample 
size, leading to narrower confidence intervals. For both the propensity score and outcome models we include all the covariates in Table \ref{tab:variable_names_def}. Additionally, for the combined data set we include season as a covariate.

\subsubsection{Checks for Assumptions}\label{ass_checks}

As mentioned in Sections \ref{assumptions} and \ref{ate_methods}, AIPW relies on conditional exchangeability in order to create balanced groups. Figure 3
is a Love plot created in R for the combined sample using the cobalt package \citep{cobalt}, which compares the standardized mean differences between the treatment and control groups of the various covariates before and after AIPW. In the original sample, eight of the covariates have standardized differences that fall outside of the rule of thumb threshold of 0.05, including the time left variable with a value of 0.74. This indicates that the groups of TFO attempts and non-attempts vary considerably, particularly in the amount of time left in the period. This is an expected feature of the data and highlights why causal methods are necessary.

\begin{figure}[t!]
\centering
\includegraphics[width=\textwidth]{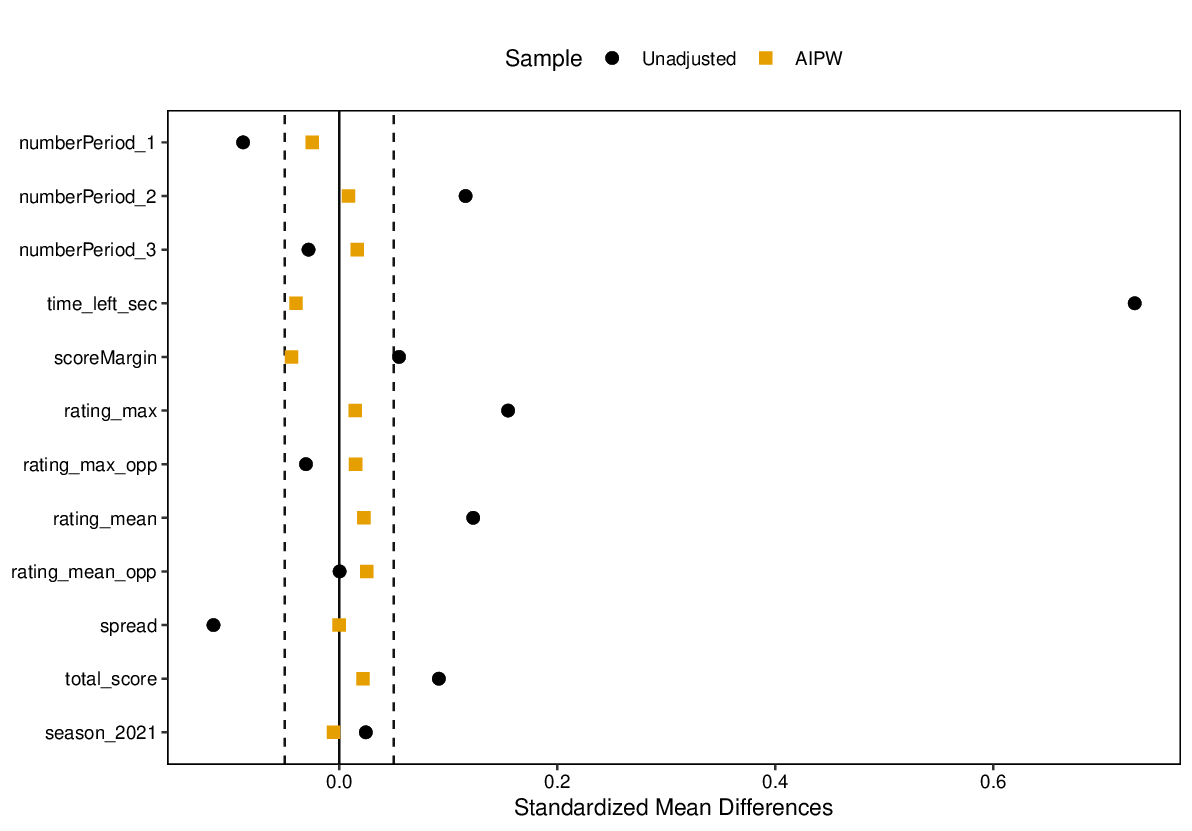} 
\caption{Standardized mean differences (treatment-control) of the covariates before and after AIPW. The dotted line represents the rule of thumb threshold of 0.05.}
\label{fig:love}
\end{figure}

Conversely, after weighting the observations, the standardized mean differences are much closer to 0, with all covariates falling inside the threshold of 0.05.  Love plots for the individual seasons (not pictured) show a similar pattern to the combined data set. Thus the weighting procedure has created groups that are comparable in terms of the measured confounders, providing evidence that the conditional exchangeability assumption is reasonable assuming the causal model represented by our DAG. In Section \ref{ate_sens} we address the possibility of unmeasured confounders with a sensitivity analysis.

The positivity assumption can also be assessed after fitting the AIPW model. Here we do so by looking at the distribution of the estimated propensity scores. Figure 4 gives histograms of the propensity scores for both groups. Other than perhaps a couple of control observations with the smallest estimated propensity scores, there is complete overlap between the two distributions. Thus we can say that the positivity assumption is reasonable for the combined data set. Similar plots for the individual seasons (not pictured) reveal that the positivity assumption is reasonable in those analyses as well. 

\begin{figure}[t]
\centering
\includegraphics[width=\textwidth]{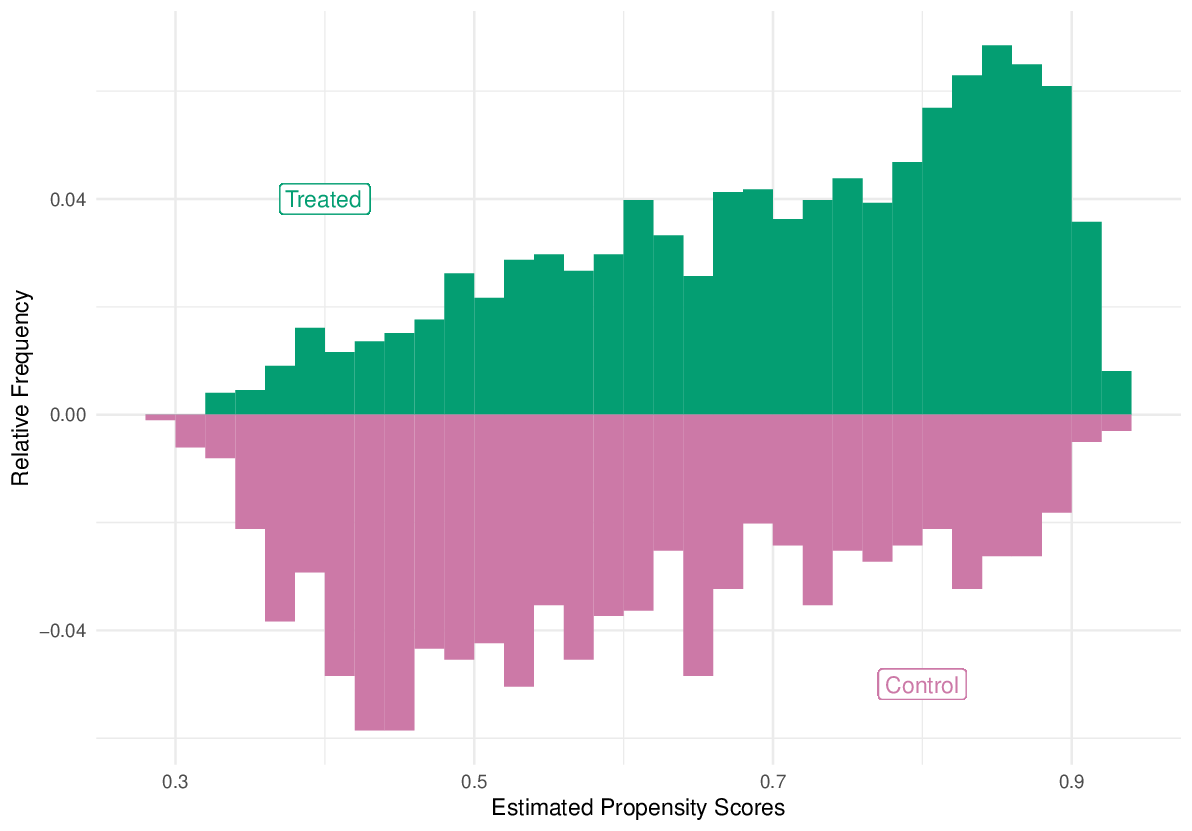} 
\caption{Density of the estimated propensity scores for treatment and control groups.}
\label{fig:density}
\end{figure}

\subsubsection{Treatment Effect Estimates}
\begin{table}[ht] 
    \centering
    \begin{tabular}{|c|c|c|c|} 
        \hline
        \multirow{3}{*}{Season} & \multicolumn{3}{c|}{Average Treatment Effect} \\ \cline{2-4} 
        & Estimate & 95\% Confidence Interval & p-value \\ \hline
        18-19 & 0.55 & (0.31, 0.78) & $<0.001$\\ \cline{1-4} 
        21-22 & 0.77 & (0.53, 1.00) & $<0.001$\\ \cline{1-4} 
        Both  & 0.66 & (0.49, 0.83) & $<0.001$  \\ \hline
    \end{tabular}
    \caption{ATE estimates, 95\% confidence intervals, and p-values for a test of a non-zero effect based on AIPW for both the individual and combined seasons.}
    \label{table:ate}
\end{table}

 Table \ref{table:ate} gives the treatment effect estimates and 95\% confidence intervals for both individual seasons and the combined data set. We estimate that if teams attempted all their TFO opportunities their average POD would be about 0.66 points higher than if they did not attempt any of their TFO opportunities, based on the combined data. A 95\% confidence interval for this effect is (0.49, 0.83). The 2021-2022 ATE estimate of 0.77 is slightly higher than the 2018-2019 estimate of 0.55, but there is considerable overlap in their confidence intervals, suggesting that the variation across individual seasons may not be significant. Overall, the positive ATE across both seasons indicates that the TFO 
strategy consistently increases the points 
differential by the end of the period. The slightly more than half a point gain may seem like a minor advantage, but NBA teams are often willing to spend effort and resources in pursuit of even relatively small edges. 

\subsubsection{Sensitivity to Unmeasured Confounding}\label{ate_sens}

While we have utilized a reasonable set of covariates based on the available data, there exists the possibility of unmeasured confounding, a challenge to many causal analyses. It is natural to ask the question of how different our estimated treatment effects might be in the presence of such unmeasured confounding. \citet{zhaosensitivity} utilize a marginal sensitivity model to construct bootstrap confidence intervals for AIPW estimators based on a parameter $\Lambda$ that measures the degree of unmeasured confounding. Specifically, in the case of a single binary unmeasured confounder $U$, the odds of an individual with $U=1$ being in the treatment group is $\Lambda$ times higher than the odds of an individual with $U=0$ being in the treatment group. As a concrete example, suppose that immediately upon his team gaining an opportunity, a coach gives a recommendation to his players about whether or not to attempt the TFO. This could certainly have an effect on whether or not the players actually attempt the TFO. Suppose that situations in which the coach recommends the attempt have $\Lambda$ times higher odds to end up as an attempt than situations in which the coach does not recommend an attempt.

Figure 5 gives 95\% confidence intervals for the ATE of a TFO attempt on the POD from the combined data set as a function of $\Lambda$ values ranging from 1.05 to 1.5, calculated using the bootsens R package \citep{bootsens}. \replaced{Note that the intervals are entirely above 0 for values of $\Lambda$ up to 1.35, but contain 0 for $\Lambda \geq 1.4$ Using our example, this would mean that even i}{I}f the odds of attempting a TFO increase by up to 35\% when a coach recommends an attempt, the treatment effect would still be statistically significant for a two-sided test at a significance level of $\alpha=.05$. However, if those odds increase by more than 40\%, the effect would no longer be statistically significant. For the 2018-2019 and 2021-2022 seasons, respectively, the treatment effect would still be statistically significant for $\Lambda$ values of up to 1.2 and 1.4, respectively. 

\begin{figure}[t!]
\centering
\includegraphics[width=\textwidth]{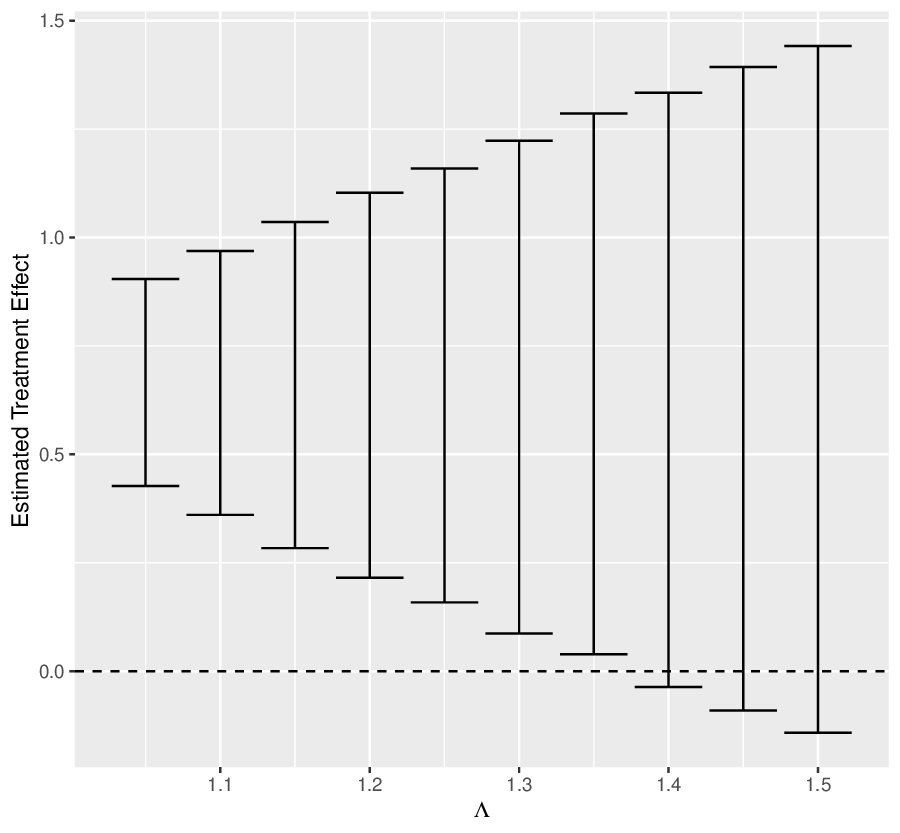} 
\caption{95\% confidence intervals for the ATE given varying degrees of unmeasured confounding indicated by $\Lambda$.}
\label{fig:sens}
\end{figure}

\added{For a more concrete example, consider the TFO attempt in the second period of the game between the Warriors and the Suns mentioned in Section \ref{example}. Our model estimated the propensity score for that TFO opportunity at 0.6654, which would imply the odds of attempting the TFO were around 2. If the true propensity score, given an unmeasured confounder, was 0.7, the odds would be 2.33 which would be 1.17 times higher than our estimate. However, if the true propensity score, given an unmeasured confounder, was 0.75, the odds would be 3 which would be 1.5 times higher than our estimate. In both individual seasons as well as the combined season, an overall $\Lambda=1.17$ would not be enough to change our conclusion, but an overall $\Lambda=2$ would.} 

Thus our results are somewhat sensitive to unmeasured confounding, particularly for the 2018-2019 season. However, in this regard having a treatment definition that relies on actions, rather than intent, may be beneficial, as it may lessen the strength of any potential confounding relationship. In our hypothetical example, due to the complexities of an NBA possession and the general autonomy of NBA players, the percentage of attempts out of all TFO opportunities may not be substantially different based on the coach's recommendation or other unmeasured confounders.

\subsubsection{Sensitivity to Treatment Definition} \label{treatsens}

As mentioned in Section \ref{definitions}, another potential concern with our analysis is the somewhat arbitrary nature of our treatment definition. We chose this definition based on the structure of NBA games, before looking at any data. However, we allow that other analysts or NBA personnel might argue for slightly different cutoffs in the TFO definition. To investigate the robustness of our treatment effect estimate to these differences, we estimate the ATE of the TFO strategy for the combined data set, along with 95\% confidence intervals, for a variety of other possible cutoffs. These results are presented in Figure 6. The horizontal axis represents the beginning and end of the TFO opportunity window, while the panels represent the cutoff for a TFO attempt. Thus the left-most category, \replaced{38}{40}-30, in the bottom \deleted{(29) }panel represents a definition in which a team needs to gain possession with between \replaced{38}{40} and 30 seconds left in the period to gain a TFO opportunity, and must take a shot or get fouled with at least 29 seconds left in the period for the opportunity to be considered an attempt. The vertical axis is the estimated ATE. The red circle represents the original definition with cutoffs of 43, 35, and 28 seconds, respectively.

\begin{figure}[t!]
\centering
\includegraphics[width=\textwidth]{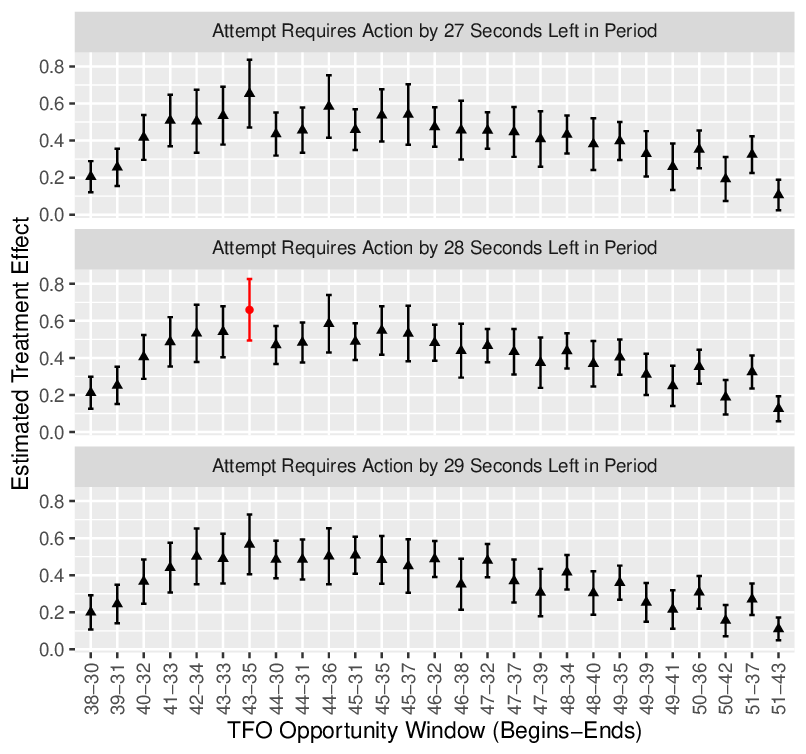} 
\caption{Treatment effect estimates and 95\% confidence intervals for alternative treatment definitions.}
\label{fig:treatsensv2}
\end{figure}

The treatment effect estimate using our original definition is the largest among the combinations of cutoffs we tested. However, there is considerable overlap between the original confidence interval and those with cutoffs fairly similar to our original definition. Note that definitions represented by the extreme left and right sides of the figure are, in our opinion, much less realistic and much less likely to be supported by other analysts or NBA personnel. For a broad swath of reasonable cutoffs, the estimated treatment effect is between 0.4 and 0.6. Thus while the treatment effect may be slightly higher with our definition than with others, the practical impact on policy may be minimal.

\subsubsection{Assessing Fourth Period Data Exclusion} \label{perperiod}

\added{In Section \ref{data}, we motivated the exclusion of fourth period and overtime possessions due to end-of-game strategic differences. This subsection provides empirical evidence supporting that choice. Table \ref{table:ateperperiod} reports average treatment effect estimates by game period for the 2021–2022 season. The first three periods yield similar estimates, whereas the estimate for the fourth period and overtime is substantially attenuated and statistically indistinguishable from zero. This pattern suggests that end-of-game possessions follow meaningfully different strategic dynamics than those in earlier periods, justifying their exclusion from our main analysis.}

\begin{table}[ht] 
    \centering
    \begin{tabular}{|c|c|c|c|} 
        \hline
        \multirow{3}{*}{Period} & \multicolumn{3}{c|}{Average Treatment Effect} \\ \cline{2-4} 
        & Estimate & 95\% Confidence Interval & p-value \\ \hline
        1 & 0.81 & (0.39, 1.23) & $<0.001$\\ \cline{1-4} 
        2 & 0.76 & (0.31, 1.20) & $<0.001$\\ \cline{1-4} 
        3 & 0.60 & (0.20, 1.00) & $0.004$  \\ \cline{1-4} 
        4/OT & 0.12 & (-0.22, 0.46) & $0.463$  \\ \hline
    \end{tabular}
    \caption{ATE estimates, 95\% confidence intervals, and p-values for a test of a non-zero effect based on AIPW for the 2021-2022 season. These effects are conditional on the period, with the fourth period and overtime periods combined.}
    \label{table:ateperperiod}
\end{table}

\subsection{Heterogeneous Treatment Effect}

\added{The average treatment effect indicates that the TFO strategy is beneficial on average, but its value may vary across game contexts. To explore when the strategy may be more or less effective, we examine heterogeneous treatment effects (HTE) using causal forests.}

\added{We augment the dataset with three additional variables—the differences in the teams’ maximum and median ratings, and the absolute score margin—which capture relative team strength and game competitiveness. We first fit a causal forest using all variables, compute variable importance, and retain those accounting for 95\% of total importance (excluding only period indicators and absolute score margin). As a check, the causal-forest ATE for the combined data set is 0.61 (SE 0.11), consistent with AIPW estimates.}

\deleted{The average treatment effect provides valuable insight, indicating that the TFO strategy is beneficial on average. However, there is a need to explore how this effect varies across different values of the covariates. Understanding this heterogeneity can help identify specific scenarios where the TFO strategy should or should not be attempted, thereby informing team strategy.}

\deleted{To this end, we investigate treatment effect heterogeneity to address the following key questions. First, is there significant heterogeneity in the treatment effect? Second, which variables contribute to this heterogeneity? We employ the causal forest framework to answer these questions.}

\deleted{Before fitting the causal forest model, we 
augment the dataset with three additional 
variables: the difference in the maximum 
ratings between the two teams, the difference 
in the median ratings between the two teams, 
and the absolute value of the score margin. 
These new variables capture 
relative differences in team strength and game 
competitiveness. Such relative measures can reveal 
heterogeneity in treatment effects not fully 
explained by absolute measures alone.
Initially, we run the causal forest model with 
all available variables. We then compute the 
variable importance measures and retain only 
those variables that cumulatively account for 
95\% of the total importance, eliminating three 
variables in the process. Variable importance was computed using the grf causal forest measure, which weights variables by split frequency and depth. We retained variables accounting for 95\% of total importance, which excluded only the period indicators and the absolute score margin. This step helps 
control the standard error.}

\deleted{As an initial check, we estimate the ATE using 
the causal forest. The estimated ATE is 0.61, 
with a standard error of 0.11, relatively close to the results obtained from AIPW. This analysis sets the foundation for a deeper 
exploration of heterogeneity in treatment 
effects, guiding future strategy development.}

\subsubsection{Overall Heterogeneity}

To begin our analysis, we first assess whether there is any evidence of treatment effect heterogeneity in the data. This is done through two separate tests.

\added{First, the test calibration procedure evaluates the null of no treatment effect heterogeneity. Once the causal forest is trained, heterogeneity is assessed via a Calibration Test \citep{chernozhukov2018generic}, which regresses the outcome on the mean forest prediction (ATE) and the out-of-bag differential forest prediction (CATE). As shown in Table \ref{table:hte_testcalibration}, the differential forest prediction yields a p-value of 0.17, providing no significant evidence against the null.}

\deleted{The first test we employ is the \emph{test 
calibration} test, which evaluates the null 
hypothesis that there is no heterogeneity in 
the treatment effect. Table \ref{table:hte_testcalibration} shows the results of the test. The p-value of the differential forest prediction 
is 0.17, indicating that we do not have 
sufficient evidence to reject the null 
hypothesis. It is important to note that the 
standard error of this test is relatively high, 
which contributes to the high p-value and 
weakens our ability to detect heterogeneity, if 
present.}

\begin{table}[h!]
\centering
\begin{tabular}{lcccc}
\toprule
 & Estimate & Std Error & t Value & p Value \\
\midrule
Mean Forest Prediction & 1.00511 & 0.17876 & 5.6227 & $<0.001$ \\
Differential Forest Prediction & 0.63352 & 0.67520 & 0.9383 & 0.1741 \\
\bottomrule
\end{tabular}
\caption{Summary of Test Calibration. The test 
for heterogeneity in treatment effects, based on 
the differential forest prediction, yields a p-
value greater than 0.1. This indicates 
insufficient evidence to reject the null 
hypothesis of no heterogeneity in treatment 
effects.}
\label{table:hte_testcalibration}
\end{table}

\added{As a complementary approach, we also compute the RATE statistic \citep{yadlowsky2024evaluating} RATE evaluates whether a prioritization rule based on predicted CATEs improves outcomes when treatment is selectively applied. We train the causal forest on 2018 data, predict CATEs for 2021, and then form the prioritization rule. The resulting RATE estimate is 0.0408 with a standard error of 0.1176, obtained via bootstrap \citep{efron1982jackknife} with 1000 samples. The confidence interval includes zero, indicating that the prioritization rule does not identify subgroups with reliably higher gains from treatment. This aligns with the test calibration results, reinforcing the conclusion that we lack strong statistical evidence of heterogeneity.}

\deleted{The second test we conduct is based on the 
Rank Average Treatment Effect (RATE)
(Yadlowsky et al., 2025). In 
this test, we define a prioritization rule and 
assess the impact of applying the treatment to 
only the top observations based on this rule. 
To implement this, we train the causal forest 
using 2018 data, predict the heterogeneous 
treatment effects for 2021 observations, and 
then use these predictions to form our 
prioritization rule. The prioritization rule is based on predicted CATEs from a causal forest trained on 2018 data and applied to 2021 data, with higher CATE predictions receiving higher priority. The RATE estimate is 
0.035 with a standard error of 0.112. The standard error is estimated using a bootstrap 
method (Efron 1982) with 200 samples. Similar 
to the first test, this result 
suggests that there is insufficient evidence to 
reject the null hypothesis of no heterogeneity in 
the treatment effect. Figure 7 shows the overall 
TOC where we can see looking at the confidence 
band that $0$ is in the confidence band.}

\deleted{(Figure 7 goes here.)}

\deleted{In summary, both the test calibration and RATE results indicate that, based on our tests, we do not find strong evidence of treatment effect heterogeneity.}

\subsubsection{Heterogeneity Across Variables}
\added{Although global tests suggest limited evidence of heterogeneity, it is still possible that treatment effects vary systematically with respect to specific variables. Prior work \citep{debosscher} highlights settings in which overall tests fail to detect heterogeneity, yet meaningful variation appears along particular covariates. Motivated by this, we examine two measures of relative team strength:
the difference in maximum player ratings between teams and the difference in median player ratings. These variables are natural candidates for effect moderation, as differences in team quality could plausibly influence the payoff of attempting a TFO.}

\added{We again rely on the RATE framework, constructing prioritization rules based solely on each variable’s value. This allows us to test whether games featuring larger rating differences exhibit systematically different treatment effects. Table \ref{tab:rate_diffs} summarizes the RATE estimates. Both are small in magnitude, with standard errors of comparable size, and neither is statistically significant. This suggests that even along these targeted axes—where we might reasonably expect strategic differences—the data do not provide evidence that the effectiveness of the TFO strategy changes meaningfully.  Thus, we conclude that there is no substantial evidence for
heterogeneity in the treatment effect across the investigated variables.}

\deleted{In the previous section, we observed that at an 
aggregate level, the tests suggest minimal 
evidence for heterogeneity in the treatment 
effect. However, prior studies such as  (De Bosscher et al., 2009) have demonstrated 
scenarios where overall homogeneity does not 
preclude the presence of heterogeneity across 
specific variables. To investigate this 
possibility, we examine heterogeneity across the 
following variables: difference in ratings of the two teams based on max rating, and difference in ratings of the two teams based on mean rating.}

\deleted{To evaluate potential heterogeneity, we employ 
the Rank Average Treatment Effect (RATE) 
framework. This method constructs prioritization 
rules based on the value of the selected variable 
and computes RATE estimates, reflecting the 
expectation that higher values of the variable 
should correspond to improved outcomes under 
treatment.}

\deleted{Table \ref{tab:rate_diffs} presents the RATE 
estimates for the selected variables. The results 
indicate that the estimates are not statistically 
significant, providing insufficient evidence to 
conclude that treatment effect heterogeneity 
exists with respect to these variables. Thus, we 
conclude that there is no substantial evidence 
for heterogeneity in the treatment effect across 
the investigated variables.}

\begin{table}[!ht]
    \centering
    \begin{tabular}{lcc}
        \toprule
        & \textbf{Estimate} & \textbf{Standard Error} \\
        \midrule
        Rating Max Diff & 0.028 & 0.085 \\
        Rating Mean Diff & 0.0825 & 0.087 \\
        \bottomrule
    \end{tabular}
    \caption{RATE estimates for investigating treatment effect heterogeneity across selected variables. The lack of statistical significance indicates minimal evidence of heterogeneity.}
    \label{tab:rate_diffs}
\end{table}

\section{Discussion}
This study explores the effectiveness of the 
two-for-one strategy in the NBA, converting a 
real-world sports strategy into a mathematical 
framework to enable causal inference. Using play-by-play data from the 2018-19 and 2021-22 NBA 
seasons, we analyze the impact of implementing 
this strategy on game outcomes by estimating the 
average increase in the point differential if all teams presented with a TFO opportunity attempt them versus if they do not attempt them. Our findings 
indicate that attempting the TFO has a 
statistically significant positive effect, 
suggesting it can be advantageous for teams in 
real game scenarios.

The results also consider potential 
heterogeneity in the effectiveness of the two-
for-one strategy using the causal forest 
framework. While we explore overall 
heterogeneity and heterogeneity across two 
selected variables, our results do not provide 
significant evidence of differential effects 
across subsets of the data. This lack of 
observed heterogeneity may imply that the 
strategy’s impact is relatively uniform across 
various game contexts and player configurations, 
though additional factors such as small sample 
size may also limit our ability to detect 
significant variation. 

While these findings contribute valuable 
insights, there are limitations to our analysis. 
First, the dataset lacks spatial information 
regarding players’ positions on the court during 
each play, which could refine estimates of shot 
difficulty and improve causal estimates. Such 
information would enhance the understanding of 
strategic choices made by players and coaches, 
but access is constrained as spatial tracking 
data is often proprietary. Additionally, our 
insignificance in heterogeneity estimates could 
result from the limited scope of our dataset, 
which includes only two seasons. Repeating the 
analysis over a larger dataset, potentially 
spanning ten seasons or more, could provide 
deeper insights and improve the statistical 
power of our heterogeneity analysis. 

Another natural extension would be to consider leagues other than the NBA, such as the Women's National Basketball Association (WNBA) or the National Collegiate Athletic Association (NCAA). Play-by-play information is generally available for these organizations as well, and it would be informative to see if similar treatment effects are found for the TFO strategy there. 

In conclusion, while our study demonstrates the 
positive impact of the TFO strategy in 
NBA games, future work with extended datasets 
and enhanced play-by-play information could 
further validate and expand on these findings. 
By overcoming data limitations and increasing 
sample size, subsequent research may uncover new 
dimensions of heterogeneity and provide a more 
comprehensive understanding of how various 
contextual factors influence the efficacy of 
this strategy.

\bibliographystyle{agsm}\renewcommand\harvardurl{\url}
\bibliography{ref}
\end{document}